\documentclass[pss]{wiley2sp}

\usepackage{amsmath}
\usepackage{amssymb}

\newcommand{\bea}{\begin{eqnarray*}}
\newcommand{\eea}{\end{eqnarray*}}
\newcommand{\bne}{\begin{equation*}}
\newcommand{\ede}{\end{equation*}}

\newcommand{\bnen}{\begin{equation}}
\newcommand{\eden}{\end{equation}}
\newcommand{\bean}{\begin{eqnarray}}
\newcommand{\eean}{\end{eqnarray}}
\newcommand{\bna}{\begin{array}}
\newcommand{\eda}{\end{array}}

\newcommand{\f}{\frac}
\newcommand{\R}{\mathfrak{R}}
\newcommand{\ki}{k_{\rm in}}

\newcommand{\igr}[2][]{\includegraphics[#1]{#2}}

\begin{document}

\title{Catastrophe optics of caustics in single and bilayer graphene:\\ fine structure of caustics}

\titlerunning{Catastrophe optics of caustics}

\author{%
Csaba P\'eterfalvi\textsuperscript{\textsf{\bfseries \Ast, 1, 2}},
Andr\'as P\'alyi\textsuperscript{\textsf{\bfseries 1, 3}},
\'Ad\'am Ruszny\'ak\textsuperscript{\textsf{\bfseries 1}},
J\'anos Koltai\textsuperscript{\textsf{\bfseries 1}},
and J\'ozsef Cserti\textsuperscript{\textsf{\bfseries 1}}}

\authorrunning{C. P\'eterfalvi et al.}

\mail{e-mail: \textsf{peterfalvi@elte.hu}}

\institute{
\textsuperscript{1}\,Institute of Physics, E\"otv\"os University, Budapest, P\'azm\'any P{\'e}ter s{\'e}t\'any 1/A, H-1117, Hungary\\
\textsuperscript{2}\,Present address: Department of Physics, Lancaster University, LA1 4YB, UK\\
\textsuperscript{3}\,Present address: Department of Physics, University of Konstanz, D-78457, Germany}

\received{XXXX, revised XXXX, accepted XXXX}
\published{XXXX}


\abstract{\abstcol{We theoretically study the scattering of a plane wave of a ballistic electron on a circular {\em n-p} junction in single and bilayer graphene. We compare the exact wave function inside the junction to that obtained from a semiclassical formula developed in catastrophe optics. In the semiclassical picture short-wavelength electrons are treated as rays of particles that can get reflected and refracted at the {\em n-p} junction according to Snell's law with negative refraction index.}{We show that for short wavelength and close to caustics this semiclassical approximation gives good agreement with the exact results in the case of single-layer graphene. We also verify the universal scaling laws that govern the shrinking rate and intensity divergence of caustics in the semiclassical limit. It is straightforward to generalize our semiclassical method to more complex geometries, offering a way to efficiently design and model graphene-based electron-optical systems.}}

\maketitle

\section{Introduction\ \ \ }
The curves of caustics, which are envelopes of a family of rays at which the density of rays is singular, can be described using geometrical optics. However, the wave intensity in the vicinity of caustics cannot be predicted by the simplest theory where rays are endowed with amplitude and phase, and allowed to interfere where they cross: as shown by Berry and Upstill~\cite{Berry_Upstill:cikk}, such calculations fail exactly on caustics by predicting intensity divergences. They have shown that applying catastrophe theory~\cite{Thom:book_es_Arnol'd:cikk} in optical systems can offer a good solution: catastrophe optics predicts finite intensity on the caustics and gives quantitative results in the short-wavelength (or 'semiclassical`) limit. This tool enables the classification of caustics according to their structural stability. Each class has its own characteristic diffraction pattern, and the corresponding wave function can be well approximated by an integral representation in terms of a polynomial describing the class~\cite{Berry_Upstill:cikk}. Near the caustics, the so-called universal scaling laws govern the $k\rightarrow \infty$ asymptotic behavior of the wave function, where $k$ is the wave number. In particular, as $k$ increases the intensity also increases, and the diffraction fringe dimensions decrease proportionally to certain power functions of $k$ with universal exponents.

Many theoretical~\cite{Veselago:cikk_Pendry1:cikk_Pendry2:cikk_Smith:cikk} and experimental~\cite{Shelby:cikk_Houck:cikk_Parazzoli:cikk_Notomi:cikk_Luo:cikk} works have already been published on materials with negative refractive index. Recently, it has been demonstrated that in graphene-nanostructures, the optics of electron flow can be described by reflections and refractions with negative refractive index~\cite{Falko_optics:cikk}, moreover, it has also been shown that in such systems, caustics can arise~\cite{Falko_optics:cikk,Cserti:cikk,Peterfalvi2009:cikk}.

With the example of a circular {\em n-p} junction (NPJ)~\cite{Cserti:cikk,Peterfalvi2009:cikk}, we investigate in this paper whether catastrophe optics could be applied in graphene in the short wavelength limit. Our main finding is that in single-layer graphene (SLG), near the caustics the wave function can be well approximated using an integral formula developed in catastrophe optics. This offers an efficient way to design and model graphene-based electron-optical devices.

\section{Catastrophe theory and integral representation of the wave function\ \ \ }\label{int_rep}

\begin{figure}[!t]
\sidecaption
\igr[scale=0.28, trim=0 -50 30 0]{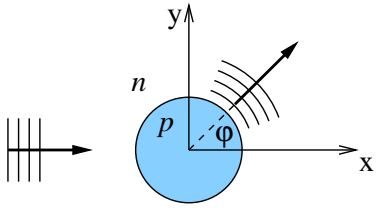}
\caption{\label{setup_abra:fig}
An incident plane wave of a ballistic electron is scattered by a circular potential barrier $V(r)$. The incoming electron on the {\em n} side is in the conduction band, while on the {\em p} side in the valence band.}
\end{figure}

Let us consider the scattering of ballistic electrons in graphene in the setup shown in Fig.~\ref{setup_abra:fig}.
We use a circular NPJ, formed by a simple gate potential $V(r) = V_0 \Theta(R-r)$, where $\Theta$ denotes the Heaviside function, $R$ is the radius of the junction and the energy $E$ of the incident particles fulfills $0<E<V_0$. The assumption of such a sharp potential is valid provided that $\lambda_F \gg d$, where $\lambda_F$ is the Fermi wavelength and $d$ is the characteristic length scale in which the scattering potential varies. To prevent intervalley scattering, we assume that $d \gg a$, where  $a$ is the lattice constant of graphene~\cite{Falko_optics:cikk,Katsnelson-Klein:cikk}. The exact calculations of the wave function inside and around the junction are presented in Ref.~\cite{Cserti:cikk} for SLG and in Ref.~\cite{Peterfalvi2009:cikk} for bilayer graphene (BLG). In the same papers it has been shown that caustics clearly arise in the wave function pattern. This suggests that we consider the propagation of electrons as that of rays of light, and caustics as the envelopes of these rays. These can be refracted and reflected according to Snell's law:
\bnen
\f{\sin \alpha}{\sin \beta}\ =\ -\f{\ki}{k}\ \equiv\ n\ ,\label{snells:eq}
\eden
where $\alpha$ and $\beta$ are the angles of incidence and refraction, respectively, and $\ki$ and $k$ are the wave numbers inside and outside the junction. Just like in certain photonic crystals~\cite{plano-concave:cikk}, in graphene NPJs, the refractive index $n$ is negative as shown in Ref.~\cite{Falko_optics:cikk}. The curves of the caustics can be calculated using differential geometry, and the results are given by Eq.~(9) in Ref.~\cite{Cserti:cikk}. Note that the caustics' curves, which are the same in SLG and in BLG, depend only on $n$. For any given $n$, we can label the curves of the caustics according to the number of internal reflections of the rays that create them. The first order caustic ($p=1$) corresponds to the envelope of the rays that were not reflected inside the junction, and the second order caustic ($p=2$) is formed by the rays after one reflection. Thus $p-1$ gives the number of the internal reflections.

We now consider the problem in terms of catastrophe optics. As catastrophe theory~\cite{Thom:book_es_Arnol'd:cikk} is a special case of the more general singularity theory, catastrophe optics deals with singularities of ray families called caustics (for details see, e.g., Ref.~\cite{Berry_Upstill:cikk}).

\begin{figure}[!t]
\sidecaption
\igr[scale=0.44, trim=0 0 0 0]{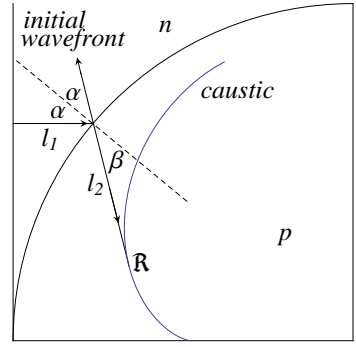}
\caption{\label{caustics:fig}
An incoming ray of electrons partly reflects, and partly enters the NPJ. The upper branch of the $p=1$ fold caustic is shown in the 2nd quarter of the NPJ. The dashed line is normal to the junction. $\R$ denotes the observation point.\newline \newline}
\end{figure}

Let us denote the optical length by $\phi$, which is the path integral weighted by the position-dependent refraction index $n$: \mbox{$\phi(\alpha,\R)=l_1+nl_2$}, for the notation see Fig.~\ref{caustics:fig}. To find the extrema of the action, we have the following condition for any given $\R$: $\partial \phi(\alpha,\R) / \partial \alpha =0$. This results in a set of $\alpha$ corresponding to the rays passing through the point $\R$. Basically, this is the well-known Fermat's principle. Varying $\R$, the extremum points $\alpha(\R)$ will change. When two or more extrema coalesce, $\phi$ will be stationary to higher than first order. It means $\R$ is on a caustic, if and only if \ $\partial^2 \! \phi(\alpha,\R) / \partial \alpha^2 =~0$. For example, on a fold type caustic, two rays touch, while in a cusp type, three.

For short wavelength, the stationary phase approximation can be used, and one can simply sum up the contributions of the rays passing through $\R$. The problem is that this sum diverges exactly on caustics where, however, the integral representation (IR) of the wave function behaves well, and gives a smooth and accurate solution:
\bnen
\psi_\textrm{ir}(\R) = c \sqrt{k} \int_{-\gamma}^{\gamma} \mathrm{d} \alpha \ b(\alpha) \exp(i k \phi(\alpha,\R))\ ,
\label{int_rep:eq}
\eden
where $b(\alpha)$ is a weighting function, $c$ is a constant prefactor and $\gamma=\arcsin(\min(1,|n|))$. Although the general form of Eq.~(\ref{int_rep:eq}) was originally derived from the Helmholtz equation~\cite{Berry_Upstill:cikk}, it can be used in SLG because the SLG eigenstates solve the Helmholtz equation. If we neglect the evanescent waves, the BLG eigenstates are solutions as well. However the evanescent waves cannot be neglected close to the boundary of the NPJ, and so we cannot expect that Eq.~(\ref{int_rep:eq}) would give an accurate result in BLG not even far from the boundary. Note that in graphene systems, $b(\alpha)$ is a spinor and it can be determined from the boundary conditions following the derivation presented in Ref.~\cite{Berry_Upstill:cikk}. We determined the $c$ prefactor by fitting $\psi_\textrm{ir}(\R)$ to the exact results.

The reason for the failure of the stationary phase approximation is that the ray contributions to the wave function are always considered separately. It is clear from Eq.~(\ref{int_rep:eq}) that for the limit $k \rightarrow \infty$, the integrand is a rapidly oscillating function of $\alpha$, and destructive interference occurs for every $\alpha$ except for those satisfying $\partial \phi(\alpha,\R) / \partial \alpha =0$. Therefore far from caustics, the stationary points are well separated, and the method of stationary phase can be used in its simplest form. On the other hand, near a caustic, two or more of the stationary points lie close together, and as soon as the distance between them is comparable to the wavelength of the particles, their contributions are not separated anymore by a region of destructive interference. In this case, they cannot be simply added as in the stationary phase approximation.

\vfill
\section{Numerical results\ \ \ }\label{num_res}

To check the accuracy of the integral approximation~(\ref{int_rep:eq}), we calculate the wave function in SLG and in BLG for many different $n$ and $p$ values. In each case, we compare the exact results for the probability function $|\psi(x,y)|^2$ inside the NPJ obtained by the exact calculation~\cite{Cserti:cikk,Peterfalvi2009:cikk} to the approximated ones. To make a quantitative comparison, we introduce the Frobenius norm of a matrix $A$ defined as $\sqrt{{\rm Tr}(A^\dag A)}= \sqrt{\sum_{i,j}|A_{i,j}|^2}$, where $A^\dag$ denotes the adjoint matrix of $A$. We define the error of our approximation as the ratio of the Frobenius norm of the difference matrix $|\psi_\textrm{ir}(x_i,y_j)|^2-|\psi(x_i,y_j)|^2$ and that of the matrix $|\psi(x_i,y_j)|^2$, where $x_i$ and $y_j$ are the coordinates of the grid points inside the NPJ, $|\psi(x_i,y_j)|^2$ is the exact probability function, and $|\psi_\textrm{ir}(x_i,y_j)|^2$ is the electron density obtained from Eq.~(\ref{int_rep:eq}).

\begin{figure}[t]%
\subfloat[Exact solution: $|\psi(x,y)|^2$.]{\label(SubFig1){sle_n1p1:fig}%
\igr[width=\linewidth]{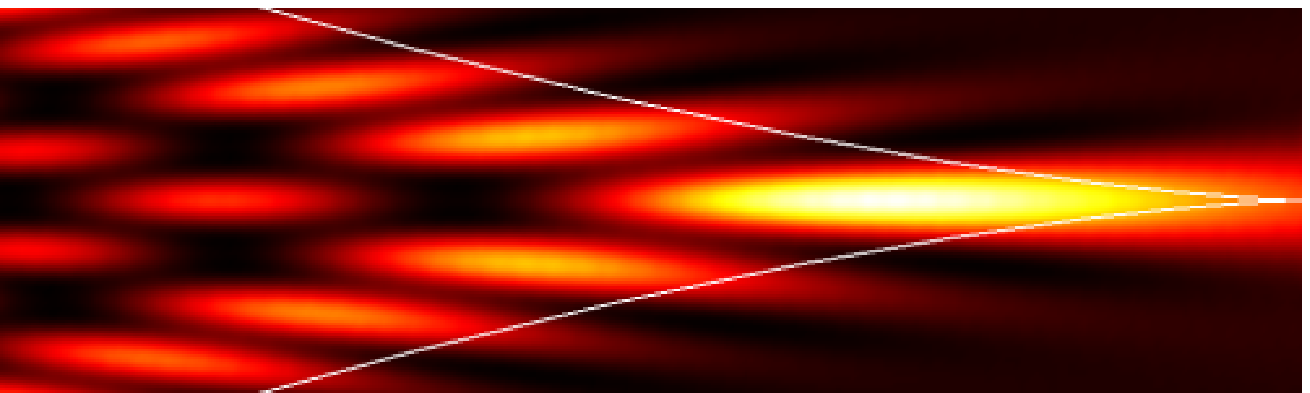}}\\
\subfloat[Semiclassical approximation: $|\psi_\textrm{ir}(x,y)|^2$.]{\label(SubFig2){slb_n1p1:fig}%
\igr[width=\linewidth]{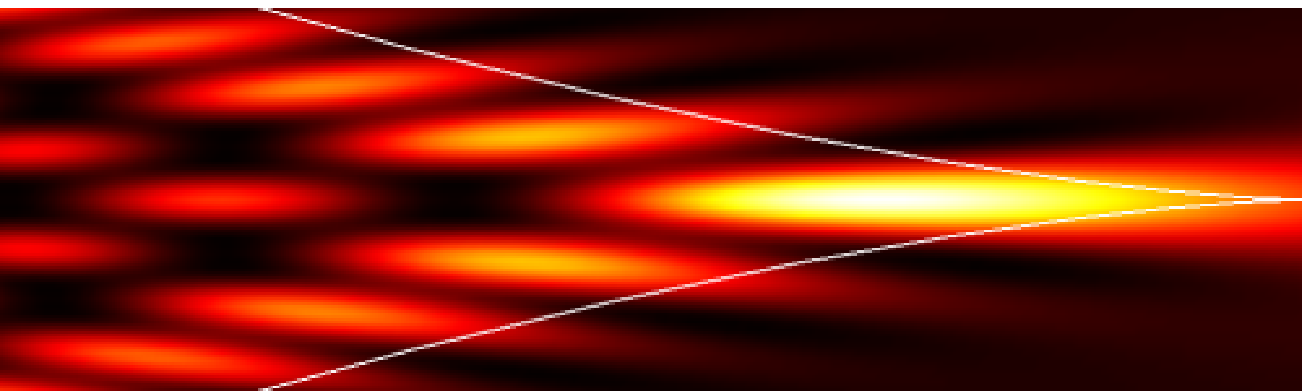}}\\
\subfloat{\igr[width=\linewidth]{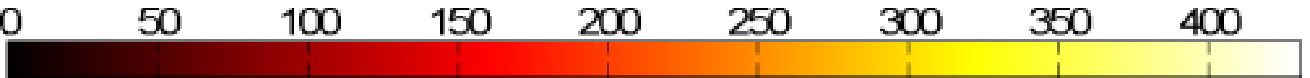}}%
\caption{\ \ Intensity around the focal point in SLG.}
\label{sl_n1p1:fig}
\end{figure}

\begin{figure}[t]%
\subfloat[Exact solution: $|\psi(x,y)|^2$.]{\label(SubFig1){ble_n1p1:fig}%
\igr[width=\linewidth]{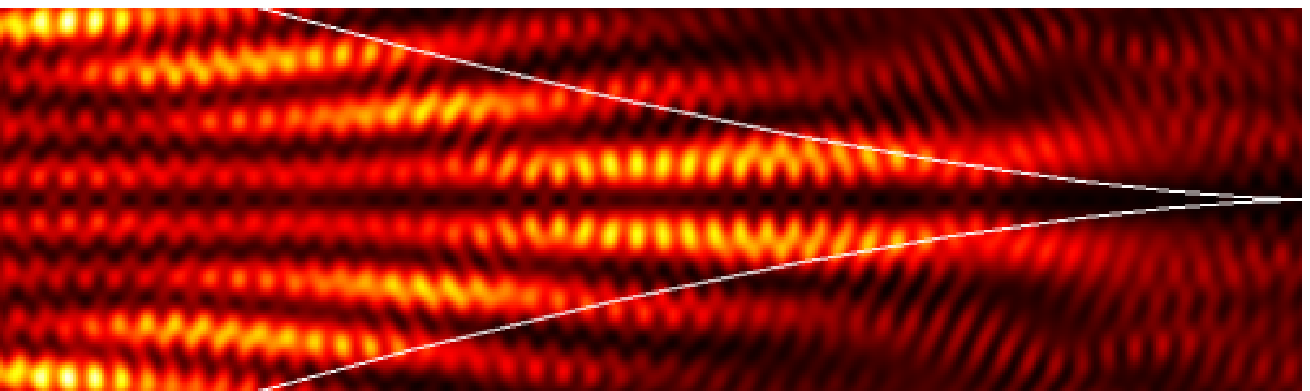}}\\
\subfloat[Semiclassical approximation: $|\psi_\textrm{ir}(x,y)|^2$.]{\label(SubFig2){blb_n1p1:fig}%
\igr[width=\linewidth]{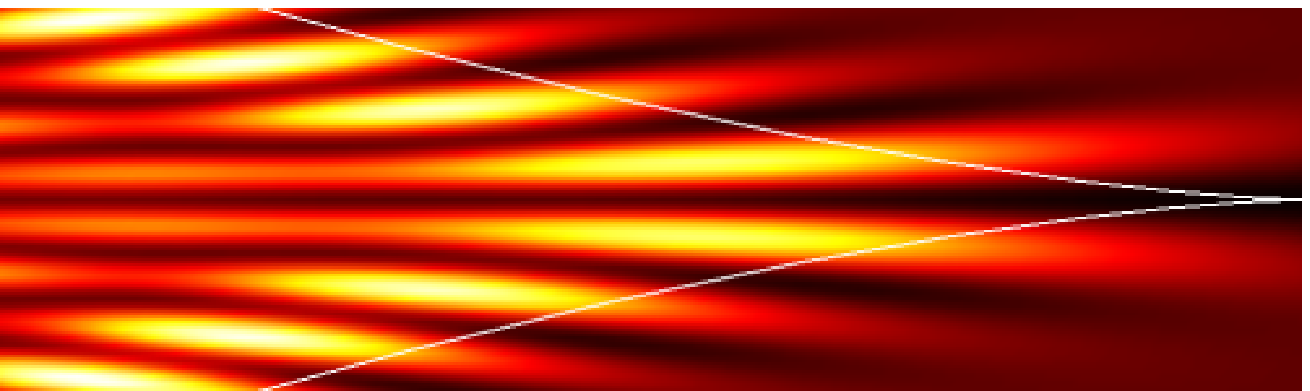}}\\
\subfloat{\igr[width=\linewidth]{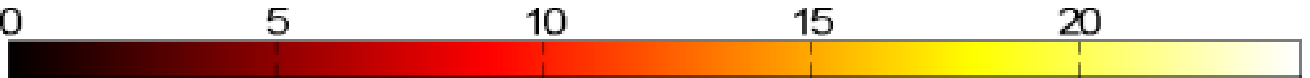}}%
\caption{\ \ Intensity around the missing focal point in BLG.}
\label{bl_n1p1:fig}
\end{figure}

Here we present only the case of $p=1$ and $n=-1$. One can clearly see in Figs.~\ref{sl_n1p1:fig} and~\ref{bl_n1p1:fig} that the approximation works well indeed. Here $k R=\ki R= 10000$, the plotted range is $-0.517<x<-0.5$, $-0.0025<y<0.0025$, and the solid line shows the analytic curve~\cite{Cserti:cikk}. Note that in BLG the focus is missing due to the zero transmission probability at perpendicular incidence~\cite{Peterfalvi2009:cikk}.
We calculate the error as discussed above, and we find that it is $0.01089$ in SLG, and $0.2038$ in BLG. The larger error for BLG is due to an apparent deviation from the exact results, namely the undulation in Fig.~\ref{ble_n1p1:fig}, which is missing in Fig.~\ref{blb_n1p1:fig}. Note that the undulations wavelength is in the order of the particles' de Broglie wavelength.

\begin{figure}[!h]
\igr[width=\linewidth]{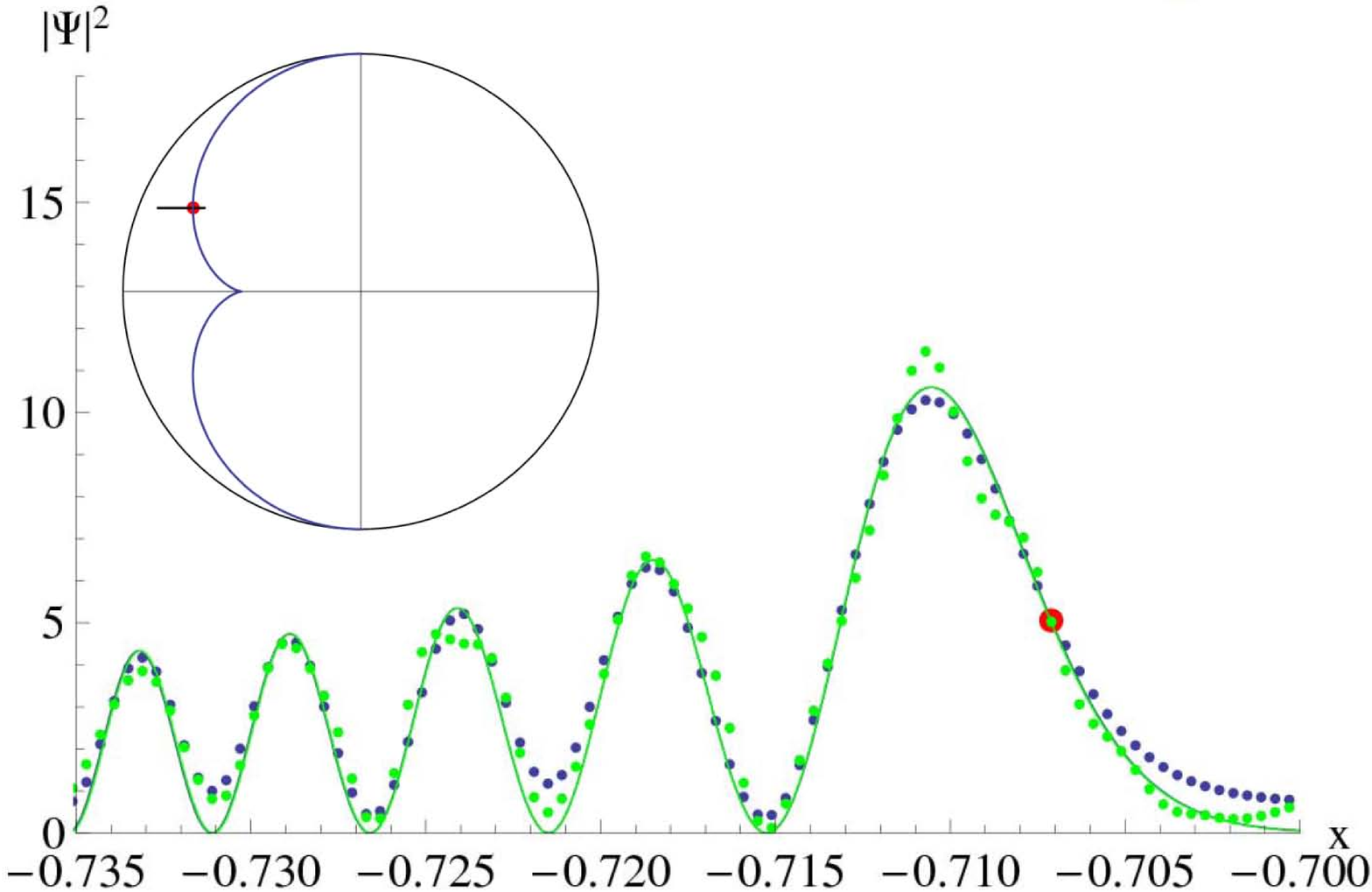}
\caption{\label{airy:fig}
Intensity in SLG as the function of $x$ close to the fold caustics by $k R=\ki R= 2500$, $n=-1$, and $p=1$. Green (bright) points indicate the exact results, and blue (dark) ones the approximation. One red (large) point shows the crossing point of the segment and the fold caustic at $(-1/\sqrt{2},1/\sqrt{8})$. The fitted $a {\rm Ai}^2(bx+c)$ type functions completely overlap for the exact and approximated results (solid line). (Here $a,b,$ and $c$ are the fitting parameters of the Airy function.)}
\end{figure}

In Fig.~\ref{airy:fig} we present results for the fold type caustics. In SLG, the intensity of the wave function is calculated along a segment parallel to the $x$-axis passing through a point on the fold caustic, at which the tangent of the caustic is parallel to the $y$-axis (see inset). According to the theory outlined in Ref.~\cite{Berry_Upstill:cikk}, the wave function along this segment should be proportional to the Airy function. Here we investigate whether the intensity follows the squared Airy function. One can see that the two curves, one fitted on the exact and the other on the approximated intensities, are basically the same, they cannot be distinguished.\\

\vfill
\subsection{Scaling laws\ \ \ }\label{sclaws:subsec}

Catastrophe theory predicts certain scaling laws that govern the $k\rightarrow \infty$ asymptotic behavior of the wave function, i.e., the shrinking rate of the diffraction fringes in the different dimensions and the divergence of the intensity. Near the caustics, the scaling laws relate the $k$-independent standard diffraction catastrophe $\Psi(\R)$~\cite{Berry_Upstill:cikk} to the $k$-dependent physical wave function $\psi_\textrm{ir}(\R)$ given by Eq.~(\ref{int_rep:eq}):
\bne
\psi_\textrm{ir}(r_i) = \kappa^{\beta} \Psi(\kappa^{\sigma_i}r_i),
\ede
where $r_i$ are the coordinates set to the caustics in the $i$th dimension, and $\kappa=k R$ is the dimensionless wave number. Here $\beta$ is the {\it singularity index}, the measure of the divergence at the most singular point, where $r_i=0$ for all $i$, and $\sigma_i$ are related to the fringe spacings in the different control directions. These indices are invariant for each classes: $\beta=1/6$ and $\sigma_1=2/3$ for fold catastrophes, while $\beta=1/4$, $\sigma_1=1/2$, and $\sigma_2=3/4$ for cusp type caustics. Note that the shrinking is anisotropic since $\sigma_1\neq \sigma_2$. Around the cusp type focal point in the NPJ, the coordinates $(r_1,r_2)$ coincide with the coordinates $(x,y)$ with the exception that its origin is in the focus. On the fold caustic, the only axis of the local coordinate system for the control parameter $r_1$ is perpendicular to the caustic, and the origin is in the crossing point on the caustic.

We have performed several numerical tests to check the scaling laws, here we present two of them. In each cases we fitted the theoretical curves on the exactly calculated results for the intensity $|\psi(x,y)|^2$. \textbf{(I)} We calculated the intensity as a function of $k$ in the focus in SLG for $p=1$ and $n=-1$. (Hereafter, the parameter $k$ is in units of $1/R$.) By fitting a straight line on $\log |\psi|$ as a function of $\log k$, for the interval $k=100...10^5$, we obtained $0.2554 \pm 0.0006$ for $\beta$. However for the interval $k=10^4...10^5$ we obtained $0.24995 \pm 0.00148$, which comes even closer to the theoretical value of $1/4$. \textbf{(II)} After choosing an arbitrary initial point close to the focus, we moved it closer to the focus step-by-step by increasing $k$ and calculating its new position from the old coordinates with the scaling laws for the fringe shrinking rate (SLG, $p=1$, $n=-1$). In every step, we recorded the intensity. From the fitting on data points spanning the interval $k=100...10^5$, we obtained that $\beta=0.2615 \pm 0.0010$, and for the interval $k=10^4...10^5$ we got that $\beta=0.2519 \pm 0.0006$. The agreement with the theoretical exponents improves as $k$ increases in all of the cases. Note that test (II) is also an indirect proof for the scaling laws regarding the shrinking rates. We have found good agreement for the scaling laws applied on the fold caustics as well. We note however that in BLG the wave pattern around the missing focus follows only the scaling laws regarding the shrinking rate.

We conclude that even for the values of $k$ we have performed our numerical calculations with, the resulting scaling exponents agree well with those predicted by the catastrophe theory in the limit of $k \rightarrow \infty$.

\vfill
\section{Summary\ \ \ }\label{summary:sec}

We have studied the scattering of a plane wave of electrons on a circular {\em n-p} junction in single and bilayer graphene with negative refractive index. We applied the semiclassical approximation of the integral representation of the wave function, and tested the predictions of catastrophe optics on our electron-optical system. Based on a number of numerical calculations, we have demonstrated that the exact wave functions are in good agreement with the approximated results close to the caustics in single-layer graphene. Defining a quantitative error for their deviations, we have found that the agreement improves in the short wavelength limit. We have also verified the scaling laws describing the asymptotic ($k\rightarrow\infty$) behavior of the intensity patterns in the vicinity of the caustics. We emphasize that in the case of a more complex scattering geometry, when exact analytical or numerical calculations are inaccessible, the semiclassical approach used here still offers a way to obtain the electron wave intensities at the caustics. Therefore we think that this approach might be useful for design and modelling of future graphene-based electron-optical devices.

\begin{acknowledgement}
\ \ \ We are grateful to Viktor Z\'olyomi for useful discussions. This work is supported by the Hungarian Science Foundation OTKA under the contracts No.~T48782 and~75529.
\end{acknowledgement}

\end{document}